\documentclass[showpacs,nofootinbib,showkeys,eqsecnum,prd,aps,preprint]{revtex4-1}
\usepackage{amssymb}
\usepackage{amsmath}
\usepackage{amsfonts}
\usepackage{graphicx}
\usepackage{bm}
\usepackage[T1]{fontenc}
\usepackage{graphicx}

\begin{document}

\title{\textbf{Particle creation by peak electric field}}
\author{T. C. Adorno${}^{a}$}
\email{tg.adorno@mail.tsu.ru, tg.adorno@gmail.com}
\author{S. P. Gavrilov${}^{a,c}$}
\email{gavrilovsergeyp@yahoo.com, gavrilovsp@herzen.spb.ru}
\author{D. M. Gitman${}^{a,b,d}$}
\email{gitman@if.usp.br}
\date{\today }

\begin{abstract}
The particle creation by the so-called peak electric field is considered.
The latter field is a combination of two exponential parts, one
exponentially increasing and another exponentially decreasing. We find exact
solutions of the Dirac equation with the field under consideration with
appropriate asymptotic conditions and calculate all the characteristics of
particle creation effect, in particular, differential mean numbers of
created particle, total number of created particles, and the probability for
a vacuum to remain a vacuum. Characteristic asymptotic regimes are
discussed in detail and a comparison with the pure asymptotically decaying
field is considered.
\end{abstract}

\affiliation{$^{a}$Department of Physics, Tomsk State University, 634050, Tomsk, Russia\\
$^{b}$P. N. Lebedev Physical Institute, 53 Leninskiy Prospekt, 119991,
Moscow, Russia\\
$^{c}$Department of General and Experimental Physics, Herzen State
Pedagogical University of Russia, Moyka Embankment 48, 191186
St.~Petersburg, Russia\\
$^{d}$Institute of Physics, University of S\~{a}o Paulo, CP 66318, CEP
05315-970, S\~{a}o Paulo, SP, Brazil;\\
}

\maketitle

\section{Introduction\label{S1}}

Particle creation from the vacuum by strong external electromagnetic fields
was studied already for a long time; see, for example, Refs. ~\cite%
{Sch51,Nikis70a,Nikis70b,Gitman77,Gitman77b,GMR85,FGS,BirDav82,Grib,GelTan15,
GelTan15b,GavGT06,GavGit15}. To
be observable, the effect needs very strong electric fields in magnitudes
compared with the Schwinger critical field. Nevertheless, recent progress in
laser physics allows one to hope that an experimental observation of the
effect can be possible in the near future, see Refs. \cite{Dun09,Dun09b,Dun09c,Dun09d,Dun09e} for the
review. Electron-hole pair creation from the vacuum becomes also an
observable effect in graphene and similar nanostructures in laboratory; see,
e.g., \cite{dassarma,dassarmab}. The particle creation from the vacuum by external
electric and gravitational backgrounds plays also an important role in
cosmology and astrophysics \cite{BirDav82,Grib,GelTan15,GelTan15b,AndMot14}.

It should be noted that the particle creation from the vacuum by external
fields is a nonperturbative effect and its calculation essentially depends
on the structure of the external fields. Sometimes calculations can be done
in the framework of the relativistic quantum mechanics, sometimes using
semiclassical and numerical methods (see Refs. \cite{BirDav82,Grib,GelTan15,GelTan15b} for
the review). In most interesting cases, when the semiclassical approximation
is not applicable,{\Large \ }the most convincing consideration of the effect
is formulated in the framework of quantum field theory, in particular, in
the framework of QED, see Ref. \cite{Gitman77,Gitman77b,FGS,GavGT06,GavGit15} and is
based on the existence of exact solutions of the Dirac equation with the
corresponding external field. Until now,\emph{\ }only few exactly solvable
cases are known for either time-dependent homogeneous or constant
inhomogeneous electric fields. One of them is related to the constant
uniform electric field \cite{Sch51}, another one to\ the\ so-called
adiabatic electric field $E\left( t\right) =E\cosh ^{-2}\left( t/T_{\mathrm{S%
}}\right) \,$\cite{NarNik70} (see also \cite{DunHal98,GavGit96}), the case
related to the so-called $T$-constant electric field \cite%
{BagGitShv75,GavGit96,GavGit08,GavGit08b}, the case related to a periodic alternating
electric field \cite{NarNIk74,NarNIk74b}, and several constant inhomogeneous electric
fields of the similar forms where the time $t$ is replaced by the spatial
coordinate $x$. The existence of exactly solvable cases of
particle creation is extremely important both for deep understanding of
quantum field theory in general and for studying quantum vacuum effects in
the corresponding external fields. In our recent work \cite{AdoGavGit14}, we
have presented a new exactly solvable case of particle creation in an
exponentially decreasing in time electric field.

In the present article, we consider for the first time particle creation in
the so-called peak electric field, which is a combination of two exponential
parts, one exponentially increasing and the other
exponentially decreasing. This is another new exactly solvable case. We
demonstrate that in the field under consideration, one can find exact
solutions with appropriate asymptotic conditions and perform nonperturbative
calculations of all the characteristics of particle creation process. In
some respects, the peak electric field shares similar features with the
Sauter-like electric field, while in other respects it{\Huge \ }can be
treated as a pulse created by laser beams. Switching the peak field on and
off, we can imitate electric fields that are specific to condensed matter
physics, in particular to graphene or Weyl semimetals as was reported, e.g.,
in Refs. \cite%
{lewkowicz10,lewkowicz10b,lewkowicz10c,Vandecasteele10,GavGitY12,Zub12,KliMos13,Fil+McL15,VajDorMoe15}.

In our calculations, we use the general theory of Ref. \cite{Gitman77,Gitman77b,FGS}
and follow in the main the consideration of particle creation effect in a
homogeneous electric field \cite{GavGit96}. To this end we find complete
sets of exact solutions of the Dirac and Klein-Gordon equations in the peak
electric field and use them to calculate differential mean numbers of
created particle, total number of created particles, and the probability for
a vacuum to remain a vacuum.\textrm{\ }Characteristic asymptotic regimes
(slowly varying peak field, short pulse field, and the most asymmetric case
related to exponentially decaying field) are discussed in detail and a
comparison with the pure asymptotically decaying field is considered.

\section{Peak electric field\label{S2}}

\subsection{General}

In this section we introduce the so-called peak electric field, that is a
time-dependent electric field directed along an unique direction\footnote{%
Greek indices refer to the Minkowski spacetime $\mu =0,...,D$ while Latin
indices refer to Euclidean space $i=1,...,D$. Here $d=D+1$ is the dimension
of the spacetime. Bold letters represent Euclidean vectors such as $\mathbf{%
r}=x^{1},x^{2},...,x^{D}$. The Minkowski metric tensor is diagonal $\eta
_{\mu \nu }=\mathrm{diag}\underset{d}{\underbrace{\left( +1,-1,...,-1\right)
}}$.}%
\begin{equation}
\mathbf{E}\left( t\right) =\left( E^{i}\left( t\right) =\delta
_{1}^{i}E\left( t\right) \,,\ \ i=1,...,D\right) \,,  \label{s3.1}
\end{equation}%
switched on at $t=-\infty $, and\ off at $t=+\infty ,$ its maximum $E>0$
occurring at a very sharp time instant, say at $t=0$, such that the limit%
\begin{equation}
\lim_{t\rightarrow -0}\dot{E}\left( t\right) \neq \lim_{t\rightarrow +0}\dot{%
E}\left( t\right) \,,  \label{s3.3}
\end{equation}%
is not defined. The latter property implies that a peak at $t=0$ is present.
Time-dependent electric fields of this form can, as usual in QED with
unstable vacuum \cite{GMR85,FGS,BirDav82,Grib} (see also \cite{AdoGavGit15}), be
described by $t$-electric potential steps,%
\begin{eqnarray}
&&A^{0}=0\,,\ \ \mathbf{A}\left( t\right) =\left( A^{i}\left( t\right)
=\delta _{1}^{i}A_{x}\left( t\right) \right) \,,  \notag \\
&&\dot{A}_{x}\left( t\right) =\frac{dA_{x}\left( t\right) }{dt}\leq
0\rightarrow \left\{
\begin{array}{l}
E\left( t\right) =-\dot{A}_{x}\left( t\right) \geq 0 \\
A_{x}\left( -\infty \right) >A_{x}\left( +\infty \right)%
\end{array}%
\right. \,,  \label{s3.2}
\end{eqnarray}%
where $A_{x}\left( -\infty \right) $, $A_{x}\left( +\infty \right) $ are
constants (for further discussion and details concerning the definition of $%
t $-electric potential steps; see Ref. \cite{AdoGavGit15}).

To study the peak electric field we consider an electric field that is
composed of independent parts, wherein for each one the Dirac equation is
exactly solvable. The field in consideration grows exponentially from the
infinitely remote past $t=-\infty $, reaches a maximal amplitude $E$ at $t=0$
and decreases exponentially to the infinitely remote future $t=+\infty $. We
label the exponentially increasing interval by $\mathrm{I}=\left( -\infty ,0%
\right] $ and the exponentially decreasing interval by $\mathrm{II}=\left(
0,+\infty \right) $, where the field and its $t$-electric potential step are%
\begin{equation}
E\left( t\right) =E\left\{
\begin{array}{l}
e^{k_{1}t}\,,\ \ t\in \mathrm{I}\,, \\
e^{-k_{2}t}\,,\ \ t\in \mathrm{II}%
\end{array}%
\right. \,,\ \ A_{x}\left( t\right) =E\left\{
\begin{array}{l}
k_{1}^{-1}\left( -e^{k_{1}t}+1\right) ,\ \ t\in \mathrm{I}\,, \\
k_{2}^{-1}\left( e^{-k_{2}t}-1\right) \,,\ \ t\in \mathrm{II}%
\end{array}%
\right. \,.  \label{ns4.0}
\end{equation}%
Here $E,k_{1},k_{2}$ are positive constants. The field and its potential are
depicted below in Fig. \ref{Fig1}.

\begin{figure}[th]
\includegraphics[scale=0.42]{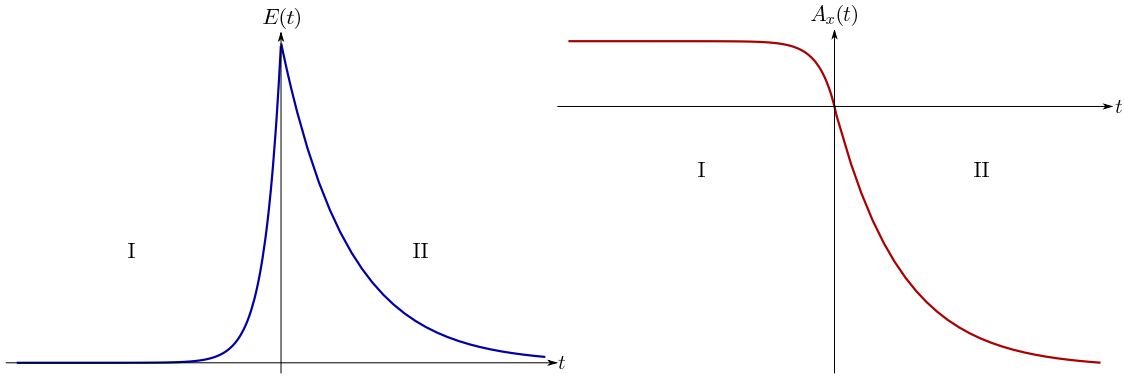}
\caption{The peak electric field $E\left( t\right) $ and its vector
potential $A_{x}\left( t\right) $ (\protect\ref{ns4.0}). Each interval is
characterized by a distinct exponential constant which explains the
non-symmetrical form of the picture. Here $k_{1}>k_{2}$ has been chosen.}
\label{Fig1}
\end{figure}

\subsection{Dirac equation with peak electric field}

To describe the problem in the framework of QED with $t$-electric potential
steps it is necessary to solve the Dirac equation for each interval
discussed above. In any of them, the Dirac equation in a $d=D+1$ dimensional
Minkowski spacetime is, in its Hamiltonian form, represented by%
\begin{eqnarray}
&&i\partial _{t}\psi \left( x\right) =H\left( t\right) \psi \left( x\right)
\,,\ \ H\left( t\right) =\gamma ^{0}\left( \boldsymbol{\gamma }\mathbf{P}%
+m\right) \,,  \notag \\
&&\,P_{x}=-i\partial _{x}-U\left( t\right) ,\ \ \mathbf{P}_{\bot }=-i%
\boldsymbol{\nabla }_{\perp },\ \ U\left( t\right) =-eA_{x}\left( t\right)
\,,  \label{2.9}
\end{eqnarray}%
where the index $\perp $ stands for spacial components perpendicular to the
electric field, $\mathbf{x}_{\perp }=\left\{ x^{2},...,x^{D}\right\} $ and $%
\mathbf{P}_{\bot }=\left( P^{2},\ldots ,P^{D}\right) $. Here $\psi (x)$ is a
$2^{[d/2]}$-component spinor ($[d/2]$ stands for the integer part of the
ratio $d/2$), $m\neq 0$ is the electron mass, $\gamma ^{\mu }$ are the $%
\gamma ${\Huge \ }matrices in $d$ dimensions , $U\left( t\right) $ is the
potential energy of one electron, and the relativistic system of units is
used throughout in this paper ($\hslash =c=1$).

As customary for $t$-electric steps \cite{AdoGavGit15}, solutions of the
Dirac equation (\ref{2.9}) have the form%
\begin{eqnarray}
&&\psi _{n}\left( x\right) =\exp \left( i\mathbf{pr}\right) \psi _{n}\left(
t\right) ,\;\ n=(\mathbf{p},\sigma )\,,  \notag \\
&&\psi _{n}\left( t\right) =\left\{ \gamma ^{0}i\partial _{t}-\gamma ^{1}
\left[ p_{x}-U\left( t\right) \right] -\boldsymbol{\gamma }\mathbf{p}_{\bot
}+m\right\} \phi _{n}(t)\,,  \label{2.10}
\end{eqnarray}%
where $\psi _{n}\left( t\right) $ and $\phi _{n}(t)$ are spinors which
depend on $t$ alone. In fact, these are states with definite momenta $%
\mathbf{p}$. Substituting Eq. (\ref{2.10}) into Dirac equation (\ref{2.9}),
we obtain a second-order differential equation for the spinor $\phi _{n}(t)$,%
\begin{equation}
\left\{ \frac{d^{2}}{dt^{2}}+\left[ p_{x}-U\left( t\right) \right] ^{2}+\pi
_{\perp }^{2}-i\gamma ^{0}\gamma ^{1}\dot{U}\left( t\right) \right\} \phi
_{n}\left( t\right) =0\,,\;\pi _{\perp }=\sqrt{\mathbf{p}_{\perp }^{2}+m^{2}}%
\,.  \label{2.11}
\end{equation}%
We separate the spinning variables by the substitution%
\begin{equation}
\phi _{n}(t)=\varphi _{n}\left( t\right) v_{\chi ,\sigma }\,,  \label{2.12}
\end{equation}%
where $v_{\chi ,\sigma }$ for $\chi =\pm 1$ and $\sigma =(\sigma _{1},\sigma
_{2},\dots ,\sigma _{\lbrack d/2]-1})$ for $\sigma _{s}=\pm 1$ is a set of
constant orthonormalized spinors, satisfying the following equations:%
\begin{equation}
\gamma ^{0}\gamma ^{1}v_{\chi ,\sigma }=\chi v_{\chi ,\sigma }\,,\ \ v_{\chi
,\sigma }^{\dag }v_{\chi ^{\prime },\sigma ^{\prime }}=\delta _{\chi ,\chi
^{\prime }}\delta _{\sigma ,\sigma ^{\prime }\,}.  \label{e2a}
\end{equation}%
The quantum numbers $\chi $ and $\sigma _{s}$ describe spin polarization and
provide a convenient parametrization of the solutions. Since in ($1+1$) and (%
$2+1$) dimensions ($d=2,3$) there are no spinning degrees of freedom, the
quantum numbers $\sigma $ are absent. In addition, in $d>3$, Eq. (\ref{2.11}) allows one to subject the constant spinors $v_{\chi ,\sigma }$ to some
supplementary conditions that, for example, can be chosen in the form%
\begin{eqnarray}
&&i\gamma ^{2s}\gamma ^{2s+1}v_{\chi ,\sigma }=\sigma _{s}v_{\chi ,\sigma }\
\mathrm{for}\ \mathrm{even\ }d\,,  \notag \\
&&i\gamma ^{2s+1}\gamma ^{2s+2}v_{\chi ,\sigma }=\sigma _{s}v_{\chi ,\sigma
}\ \mathrm{for\ odd}\ d\,,  \label{e2.5}
\end{eqnarray}%
Then the scalar functions $\varphi _{n}\left( t\right) $ have to obey the
second-order differential equation%
\begin{equation}
\left\{ \frac{d^{2}}{dt^{2}}+\left[ p_{x}-U\left( t\right) \right] ^{2}+\pi
_{\perp }^{2}-i\chi \dot{U}\left( t\right) \right\} \varphi _{n}\left(
t\right) =0\,.  \label{s2}
\end{equation}

In $d$ dimensions, for any given set of quantum numbers $\mathbf{p}$, there
exist only $J_{(d)}=2^{[d/2]-1}$ different spin states. The projection
operator inside the curly brackets in Eq. (\ref{2.10}) does not commute with
the matrix $\gamma ^{0}\gamma ^{1}$ and, consequently, transforms $\phi
_{n}^{\left( \chi \right) }(x)$ with a given $\chi $ to a linear
superposition of functions $\phi _{n}^{\left( +1\right) }(x)$ and $\phi
_{n^{\prime }}^{\left( -1\right) }(x)$ with indices $n$ and $n^{\prime }$
corresponding to the same $\mathbf{p}$. For this reason, solutions of (\ref%
{2.10}) differing only by values of $\chi $ and $\sigma _{s}$ are linearly
dependent.\textrm{\ }That is why it is enough to select a particular value
of $\chi $ to perform some specific calculations, whose choice shall be
explicitly indicated when necessary.

Exact solutions of the Dirac equation with the exponentially decreasing
electric field have been obtained by us previously in \cite{AdoGavGit14}.
Thus, using some results of the latter work, below we summarize the
structure of solutions for each interval\textbf{\ }and unify it in a single
presentation. To this aim we introduce new variables $\eta _{j}$,%
\begin{eqnarray}
&&\eta _{1}\left( t\right) =ih_{1}e^{k_{1}t}\,,\ \ \eta _{2}\left( t\right)
=ih_{2}e^{-k_{2}t}\,,  \notag \\
&&h_{j}=\frac{2eE}{k_{j}^{2}},\ \ j=1,2\,,  \label{i.0}
\end{eqnarray}%
in place of $t$ and represent the scalar functions $\varphi _{n}\left(
t\right) $ as%
\begin{eqnarray}
&&\varphi _{n}^{j}\left( t\right) =e^{-\eta _{j}/2}\eta _{j}^{\nu _{j}}%
\tilde{\varphi}^{j}\left( \eta _{j}\right) ,  \notag \\
&&\nu _{j}=\frac{i\omega _{j}}{k_{j}}\,,\ \ \omega _{j}=\sqrt{\pi
_{j}^{2}+\pi _{\perp }^{2}}\,,\ \ \pi _{j}=p_{x}-\left( -1\right) ^{j}\frac{%
eE}{k_{j}}\,,  \label{i.2}
\end{eqnarray}%
where the subscript $j$ distinguishes quantities associated to the intervals
$\mathrm{I}$ ($j=1$) and $\mathrm{II}$ ($j=2$), respectively. Then the
functions $\tilde{\varphi}^{j}\left( \eta _{j}\right) $ satisfy the
confluent hypergeometric equation \cite{BatE53},%
\begin{equation*}
\left[ \eta _{j}\frac{d^{2}}{d\eta _{j}^{2}}+\left( c_{j}-\eta _{j}\right)
\frac{d}{d\eta _{j}}-a_{j}\right] \tilde{\varphi}^{j}\left( \eta _{j}\right)
=0\,,
\end{equation*}%
whose parameters are%
\begin{equation}
c_{j}=1+2\nu _{j}\,,\ \ a_{j}=\frac{1}{2}\left( 1+\chi \right) +\left(
-1\right) ^{j}\frac{i\pi _{j}}{k_{j}}+\nu _{j}\,.  \label{i.3}
\end{equation}%
A fundamental set of solutions for the equation is composed by two linearly
independent confluent hypergeometric functions:%
\begin{equation*}
\Phi \left( a_{j},c_{j};\eta _{j}\right) \ \ \mathrm{and}\ \ \eta
_{j}^{1-c_{j}}e^{\eta _{j}}\Phi \left( 1-a_{j},2-c_{j};-\eta _{j}\right) \,,
\end{equation*}%
where%
\begin{equation}
\Phi \left( a,c;\eta \right) =1+\frac{a}{c}\frac{\eta }{1!}+\frac{a\left(
a+1\right) }{c\left( c+1\right) }\frac{\eta ^{2}}{2!}+\ldots \,.  \label{chf}
\end{equation}%
Thus the general solution of Eq.~(\ref{s2}) in the intervals $\mathrm{I}$
and $\mathrm{II}$ can be expressed as the following linear superposition:%
\begin{align}
& \varphi _{n}^{j}\left( t\right) =b_{2}^{j}y_{1}^{j}\left( \eta _{j}\right)
+b_{1}^{j}y_{2}^{j}\left( \eta _{j}\right) \,,  \notag \\
& y_{1}^{j}\left( \eta _{j}\right) =e^{-\eta _{j}/2}\eta _{j}^{\nu _{j}}\Phi
\left( a_{j},c_{j};\eta _{j}\right) \,,  \notag \\
& y_{2}^{j}\left( \eta _{j}\right) =e^{\eta _{j}/2}\eta _{j}^{-\nu _{j}}\Phi
\left( 1-a_{j},2-c_{j};-\eta _{j}\right) \,,  \label{i.3.3}
\end{align}%
with constants $b_{1}^{j}$ and $b_{2}^{j}$ being fixed by the initial
conditions. The Wronskian of the $y$ functions is%
\begin{equation}
y_{1}^{j}\left( \eta _{j}\right) \frac{d}{d\eta _{j}}y_{2}^{j}\left( \eta
_{j}\right) -y_{2}^{j}\left( \eta _{j}\right) \frac{d}{d\eta _{j}}%
y_{1}^{j}\left( \eta _{j}\right) =\frac{1-c_{j}}{\eta _{j}}\,.  \label{i.3.4}
\end{equation}

It is worth noting that the complete set of solutions for the Klein-Gordon
equation,%
\begin{equation}
\phi _{n}\left( x\right) =\exp \left( i\mathbf{pr}\right) \varphi _{n}\left(
t\right) \,.  \label{KG1}
\end{equation}%
can be obtained from the solutions above by setting $\chi =0$ in all
formulas. In this case $n=\mathbf{p}$.

With the help of the exact solutions one may write Dirac spinors throughout
the time interval $t\in \left( -\infty ,+\infty \right) $. As can be seen
from (\ref{ns4.0}), the peak electric field is switched on at the infinitely
remote past $t\rightarrow -\infty $ and switched{\Huge \ }off at the
infinitely remote future $t\rightarrow +\infty $. At these regions, the
exact solutions represent free particles,%
\begin{equation}
\ _{\zeta }\varphi _{n}\left( t\right) =\ _{\zeta }\mathcal{N}e^{-i\zeta
\omega _{1}t}\,\mathrm{\ \ if\ }\ t\rightarrow -\infty ,\ \ ^{\zeta }\varphi
_{n}\left( t\right) =\ ^{\zeta }\mathcal{N}e^{-i\zeta \omega _{2}t}\mathrm{\
\ if}\ \ t\rightarrow +\infty \,,  \label{i.4.0}
\end{equation}%
respectively, where $\omega _{1}$ denotes energy of initial particles at $%
t\rightarrow -\infty $, $\omega _{2}$ denotes energy of final particles at $%
t\rightarrow +\infty $ and $\zeta $ labels electron $\left( \zeta =+\right) $
and positron $\left( \zeta =-\right) $ states. Here$\;_{\zeta }\mathcal{N}$
and $\;^{\zeta }\mathcal{N}$ are normalization constants with respect to the
inner product\footnote{%
For a detailed explanation concerning the inner product for $t$-electric
potential steps see, e. g., Ref. \cite{AdoGavGit15}.}%
\begin{equation}
\left( \psi ,\psi ^{\prime }\right) =\int \psi ^{\dagger }\left( x\right)
\psi ^{\prime }\left( x\right) d\mathbf{r}\,,\ \ d\mathbf{r}%
=dx^{1}...dx^{D}\,,  \label{t4.0}
\end{equation}%
These constants are%
\begin{eqnarray}
&&\ _{\zeta }\mathcal{N}=\ _{\zeta }CY\,,\ \ \ ^{\zeta }\mathcal{N}=\
^{\zeta }CY\,,\ \ Y=V_{\left( d-1\right) }^{-1/2}\,,  \notag \\
&&_{\zeta }C=\left( 2\omega _{1}q_{1}^{\zeta }\right) ^{-1/2}\,,\ \ \
^{\zeta }C=\left( 2\omega _{2}q_{2}^{\zeta }\right) ^{-1/2}\,,  \notag \\
&&q_{j}^{\zeta }=\omega _{j}-\chi \zeta \pi _{j}\,,  \label{i.4.2}
\end{eqnarray}%
where $V_{\left( d-1\right) }$ is the spatial volume. By virtue of these
properties, electron (positron) states can be selected as follows:%
\begin{eqnarray}
\ _{+}\varphi _{n}\left( t\right) &=&\;_{+}\mathcal{N}\exp \left( i\pi \nu
_{1}/2\right) y_{2}^{1}\left( \eta _{1}\right) \,,\,\ _{-}\varphi _{n}\left(
t\right) =\;_{-}\mathcal{N}\exp \left( -i\pi \nu _{1}/2\right)
y_{1}^{1}\left( \eta _{1}\right) \,,\ \ t\in \mathrm{I}\,;  \notag \\
\ ^{+}\varphi _{n}\left( t\right) &=&\;^{+}\mathcal{N}\exp \left( -i\pi \nu
_{2}/2\right) y_{1}^{2}\left( \eta _{2}\right) \,,\,\ ^{-}\varphi _{n}\left(
t\right) =\;^{-}\mathcal{N}\exp \left( i\pi \nu _{2}/2\right)
y_{2}^{2}\left( \eta _{2}\right) \,,\ \ t\in \mathrm{II}\,.  \label{i.4.1}
\end{eqnarray}

\subsection{$g$ Coefficients and mean numbers of created particles\label%
{Ss2.3}}

Taking into account the complete set of exact solutions (\ref{i.3.3}), the
functions$\ \ _{-}\varphi _{n}\left( t\right) $ and $\ ^{+}\varphi
_{n}\left( t\right) $ can be presented in the form%
\begin{eqnarray}
\ \ ^{+}\varphi _{n}\left( t\right) &=&\left\{
\begin{array}{l}
g\left( _{+}|^{+}\right) \ _{+}\varphi _{n}\left( t\right) +\kappa g\left(
_{-}|^{+}\right) \ _{-}\varphi _{n}\left( t\right) \,,\ \ t\in \mathrm{I} \\
\;^{+}\mathcal{N}\exp \left( -i\pi \nu _{2}/2\right) y_{1}^{2}\left( \eta
_{2}\right) \,,\ \ \ \ \ \ \ \ \ \ \,t\in \mathrm{II}%
\end{array}%
\right. \,,  \label{i.6.1} \\
\ _{-}\varphi _{n}\left( t\right) &=&\left\{
\begin{array}{l}
\;_{-}\mathcal{N}\exp \left( -i\pi \nu _{1}/2\right) y_{1}^{1}\left( \eta
_{1}\right) \,,\ \ \ \ \ \ \ \ \ \ \ \ \,t\in \mathrm{I} \\
g\left( ^{+}|_{-}\right) \ ^{+}\varphi _{n}\left( t\right) +\kappa g\left(
^{-}|_{-}\right) \ ^{-}\varphi _{n}\left( t\right) \,,\ \ t\in \mathrm{II}%
\end{array}%
\right. \,,  \label{i.6.2}
\end{eqnarray}%
for the whole axis $t$, where the coefficients $g$\ are the diagonal matrix
elements,%
\begin{equation}
\left( \ _{\zeta ^{\prime }}\psi _{l},\ ^{\zeta }\psi _{n}\right) =\delta
_{l,n}g\left( _{\zeta ^{\prime }}|^{\zeta }\right) ,\ \ g\left( ^{\zeta
^{\prime }}|_{\zeta }\right) =g\left( _{\zeta ^{\prime }}|^{\zeta }\right)
^{\ast }.  \label{t4.5}
\end{equation}%
These coefficients satisfy the unitary relations%
\begin{equation}
\sum_{\varkappa }g\left( ^{\zeta }|_{\varkappa }\right) g\left( _{\varkappa
}|^{\zeta ^{\prime }}\right) =\sum_{\varkappa }g\left( _{\zeta }|^{\varkappa
}\right) g\left( ^{\varkappa }|_{\zeta ^{\prime }}\right) =\delta _{\zeta
,\zeta ^{\prime }}\,.  \label{3.16.1}
\end{equation}%
Here the constant $\kappa $ ($\kappa =+1$ above) allows us to cover the
Klein-Gordon case, whose details are discussed in Eqs. (\ref{KGC}) and (\ref%
{gpKG}) below.

The functions$\ _{-}\varphi _{n}\left( t\right) $ and $\ ^{+}\varphi
_{n}\left( t\right) $ and their derivatives satisfy the following continuity
conditions:
\begin{equation}
\left. \ _{-}^{+}\varphi _{n}(t)\right\vert _{t=-0}=\left. \ _{-}^{+}\varphi
_{n}(t)\right\vert _{t=+0}\,,\ \ \left. \partial _{t}\ _{-}^{+}\varphi
_{n}(t)\right\vert _{t=-0}=\left. \partial _{t}\ _{-}^{+}\varphi
_{n}(t)\right\vert _{t=+0}\,.  \label{i.7}
\end{equation}%
Using Eq.~(\ref{i.7}) and the Wronskian (\ref{i.3.4}), one can find each
coefficient $g\left( _{\zeta }|^{\zeta ^{\prime }}\right) $ and $g\left(
^{\zeta }|_{\zeta ^{\prime }}\right) $ in Eqs.~(\ref{i.6.1}) and (\ref{i.6.2}%
). For example, applying these conditions to the set (\ref{i.6.1}), the
coefficient $g\left( _{-}|^{+}\right) $ takes the form%
\begin{eqnarray}
&&g\left( _{-}|^{+}\right) =\mathcal{C}\Delta \,,\ \ \mathcal{C}=-\frac{1}{2}%
\sqrt{\frac{q_{1}^{-}}{\omega _{1}q_{2}^{+}\omega _{2}}}\exp \left[ \frac{%
i\pi }{2}\left( \nu _{1}-\nu _{2}\right) \right] \,,  \notag \\
&&\Delta =\left. \left[ k_{1}h_{1}y_{1}^{2}\left( \eta _{2}\right) \frac{d}{%
d\eta _{1}}y_{2}^{1}\left( \eta _{1}\right) +k_{2}h_{2}y_{2}^{1}\left( \eta
_{1}\right) \frac{d}{d\eta _{2}}y_{1}^{2}\left( \eta _{2}\right) \right]
\right\vert _{t=0}\,.  \label{gp}
\end{eqnarray}%
Alternatively, we obtain from the set (\ref{i.6.1})%
\begin{eqnarray}
&&g\left( ^{+}|_{-}\right) =\mathcal{C}^{\prime }\Delta ^{\prime }\,,\ \
\mathcal{C}^{\prime }=-\frac{1}{2}\sqrt{\frac{q_{2}^{+}}{\omega
_{1}q_{1}^{-}\omega _{2}}}\exp \left[ \frac{i\pi }{2}\left( \nu _{2}-\nu
_{1}\right) \right] \,,  \notag \\
&&\Delta ^{\prime }=\left\{ k_{2}h_{2}y_{1}^{1}\left( \eta _{1}\right) \frac{%
d}{d\eta _{2}}y_{2}^{2}\left( \eta _{2}\right) +k_{1}h_{1}y_{2}^{2}\left(
\eta _{2}\right) \frac{d}{d\eta _{1}}y_{1}^{1}\left( \eta _{1}\right)
\right\} _{t=0}\,.  \label{nd5}
\end{eqnarray}%
Comparing Eqs.~(\ref{gp}) and (\ref{nd5}) one can easily verify that the
symmetry under a simultaneous change $k_{1}\leftrightarrows k_{2}$ and $\pi
_{1}\leftrightarrows -\pi _{2}$ holds,%
\begin{equation}
g\left( ^{+}|_{-}\right) \leftrightarrows g\left( _{-}|^{+}\right) \,.
\label{nd6}
\end{equation}

A formal transition to the Klein-Gordon case can be performed by setting $\chi =0$
and $\kappa =-1$\ in Eqs.~(\ref{i.6.1}) and (\ref{i.6.2}), and by replacing%
{\Huge \ }the normalization factors $_{\zeta }C,\ ^{\zeta }C$ written in (%
\ref{i.4.2}) by%
\begin{equation}
\ _{\zeta }C=\left( 2\omega _{1}\right) ^{-1/2}\,,\ \ \ ^{\zeta }C=\left(
2\omega _{2}\right) ^{-1/2}\,.  \label{KGC}
\end{equation}%
After these substitutions, the coefficient $g\left( _{-}|^{+}\right) $ for
scalar particles reads%
\begin{equation}
g\left( _{-}|^{+}\right) =\mathcal{C}_{\mathrm{sc}}\left. \Delta \right\vert
_{\chi =0}\,,\ \ \mathcal{C}_{\mathrm{sc}}=\left( 4\omega _{1}\omega
_{2}\right) ^{-1/2}\exp \left[ i\pi \left( \nu _{1}-\nu _{2}\right) /2\right]
\,,  \label{gpKG}
\end{equation}%
with $\Delta $ given by Eq.~(\ref{gp}). In this case it is worth noting that
the symmetry under the simultaneous change $k_{1}\leftrightarrows k_{2}$ and
$\pi _{1}\leftrightarrows -\pi _{2}$ holds as%
\begin{equation}
g\left( ^{+}|_{-}\right) \leftrightarrows -g\left( _{-}|^{+}\right) \,.
\label{nd6b}
\end{equation}%
Note that for scalar particles\ the coefficients $g$ satisfy the unitary
relations%
\begin{equation}
\sum_{\varkappa }\varkappa g\left( ^{\zeta }|_{\varkappa }\right) g\left(
_{\varkappa }|^{\zeta ^{\prime }}\right) =\sum_{\varkappa }\varkappa g\left(
_{\zeta }|^{\varkappa }\right) g\left( ^{\varkappa }|_{\zeta ^{\prime
}}\right) =\zeta \delta _{\zeta ,\zeta ^{\prime }}\,.  \label{3.16.2}
\end{equation}

Using a unitary transformation $V$\ between the initial and final Fock spaces,
see \cite{FGS}, one finds that the differential mean number of
electron-positron pairs created from the vacuum can be expressed via the
coefficients $g$\ as%
\begin{equation}
N_{n}^{\mathrm{cr}}=\left\vert g\left( _{-}|^{+}\right) \right\vert ^{2}
\label{mN}
\end{equation}%
both for fermions and bosons. Then the total number of \ created pairs is
given by the sum%
\begin{equation}
N=\sum_{n}N_{n}^{\mathrm{cr}}=\sum_{n}\left\vert g\left( _{-}\left\vert
^{+}\right. \right) \right\vert ^{2}  \label{TN}
\end{equation}%
and the vacuum-to-vacuum transition probability reads%
\begin{equation}
P_{v}=\exp \left\{ \kappa \sum_{n}\ln \left[ 1-\kappa N_{n}^{\mathrm{cr}}%
\right] \right\} \,.  \label{vacprob}
\end{equation}

For Dirac particles, using $g\left( _{-}|^{+}\right) $ given by Eq.~(\ref{gp}%
), we find in the case under consideration%
\begin{equation}
N_{n}^{\mathrm{cr}}=\left\vert \mathcal{C}\Delta \right\vert ^{2}\,.
\label{4.0}
\end{equation}

For scalar particles, using $g\left( _{-}|^{+}\right) $ given by Eq.~(\ref%
{gpKG}) the same quantity has the form%
\begin{equation}
N_{n}^{\mathrm{cr}}=\left\vert \mathcal{C}_{\mathrm{sc}}\left. \Delta
\right\vert _{\chi =0}\right\vert ^{2}\,.  \label{4b}
\end{equation}%
It is clear that $N_{n}^{\mathrm{cr}}$ is a function of modulus squared of
transversal momentum, $\mathbf{p}_{\perp }^{2}$. It follows from Eq.~(\ref%
{nd6}) and (\ref{nd6b}), respectively, that $N_{n}^{\mathrm{cr}}$ is
invariant under the simultaneous change $k_{1}\leftrightarrows k_{2}$ and $%
\pi _{1}\leftrightarrows -\pi _{2}$ for both fermions and bosons. Then if $%
k_{1}=k_{2}$, $N_{n}^{\mathrm{cr}}$ appears to be an even function of
longitudinal momentum $p_{x}$ too.

\section{Slowly varying field\label{S4}}

\subsection{Differential quantities\label{Ss4.1}}

We are primarily interested in a strong field, when $N_{n}^{\mathrm{cr}}$
are not necessarily small in some ranges of quantum numbers and
semiclassical calculations cannot{\Huge \ }be applied. The inverse
parameters $k_{1}^{-1}$, $k_{2}^{-1}$ represent scales of time duration for
increasing and decreasing phases of the electric field. In particular, we
have a slowly varying field at small values of both $k_{1},k_{2}\rightarrow
0 $. This case can be considered as a new two-parameter regularization for a
constant electric field [additional to the known one-parameter
regularizations by the Sauter-like electric field, $E\cosh ^{-2}\left( t/T_{%
\mathrm{S}}\right) $, and the $T$-constant electric field (an electric field
which effectively acts during a sufficiently large but finite time interval $%
T$)]. Let us consider only this case, supposing that $k_{1}$ and $k_{2}$ are
sufficiently small, obeying the conditions%
\begin{equation}
\min \left( h_{1},h_{2}\right) \gg \max \left( 1,m^{2}/eE\right) \,.
\label{4.1}
\end{equation}

Let us analyze how the numbers $N_{n}^{\mathrm{cr}}$ depend on the
parameters $p_{x}$ and $\pi _{\perp }$. It can be seen from semiclassical
analysis that $N_{n}^{\mathrm{cr}}$ is exponentially small in the range of
very large $\pi _{\perp }\gtrsim \min \left(
eEk_{1}^{-1},eEk_{2}^{-1}\right) $. Then the range of fixed $\pi _{\perp }$
is of interest, and in what follows, we assume that%
\begin{equation}
\sqrt{\lambda }<K_{\bot }\,,\ \ \lambda =\frac{\pi _{\perp }^{2}}{eE}\,,
\label{4.2}
\end{equation}%
where $K_{\bot }$ is any given number satisfying the condition%
\begin{equation}
\min \left( h_{1},h_{2}\right) \gg K_{\bot }^{2}\gg \max \left(
1,m^{2}/eE\right) \,.  \label{4.3}
\end{equation}

By virtue of symmetry properties of $N_{n}^{\mathrm{cr}}$ discussed above,
one can only consider $p_{x}$ either{\Huge \ }positive or negative. Let us,
for example, consider the interval $-\infty <p_{x}\leq 0$. In this case $\pi
_{2}$ is negative and large, $-\pi _{2}\geq eE/k_{2}$, while $\pi _{1}$
varies from positive to negative values, $-\infty <\pi _{1}\leq eE/k_{1}$.
The case of large negative $\pi _{1}$, $-2\pi _{1}/k_{1}>K_{1}$, where $%
K_{1} $ is any given large number, $K_{1}\gg K_{\bot }$, is quite simple. In
this case, using the appropriate asymptotic expressions of the confluent
hypergeometric function one can see that $N_{n}^{\mathrm{cr}}$ is negligibly
small. To see this, Eq. (\ref{A10}) in Appendix \ref{Ap2} is useful in the
range $h_{1}\gtrsim -2\pi _{1}/k_{1}>K_{1}$ and the expression for large $%
c_{2}$ with fixed $a_{2}$ and $h_{2}$ and the expression for large $c_{1}$
with fixed $a_{1}-c_{1}$ and $h_{1}$, given in \cite{BatE53}, are useful in
the range $-2\pi _{1}/k_{1}\gg h_{1}$.

We expect a significant contribution in the range%
\begin{equation}
h_{1}\geq 2\pi _{1}/k_{1}>-K_{1},  \label{4.4}
\end{equation}%
that can be divided in four subranges%
\begin{eqnarray}
\mathrm{(a)} &&\;h_{1}\geq 2\pi _{1}/k_{1}>h_{1}\left[ 1-\left( \sqrt{h_{1}}%
g_{2}\right) ^{-1}\right] ,  \notag \\
\mathrm{(b)} &&\;h_{1}\left[ 1-\left( \sqrt{h_{1}}g_{2}\right) ^{-1}\right]
>2\pi _{1}/k_{1}>h_{1}\left( 1-\varepsilon \right) ,  \notag \\
\mathrm{(c)} &&\;h_{1}\left( 1-\varepsilon \right) >2\pi
_{1}/k_{1}>h_{1}/g_{1},  \notag \\
\mathrm{(d)} &&\;h_{1}/g_{1}>2\pi _{1}/k_{1}>-K_{1},  \label{4.5}
\end{eqnarray}%
where $g_{1}$, $g_{2}$, and $\varepsilon $ are any given numbers satisfying
the condition $g_{1}$ $\gg 1$, $g_{2}$ $\gg 1$, and $\varepsilon \ll 1$. $.$%
Note that $\tau _{1}=-ih_{1}/\left( 2-c_{1}\right) \approx \frac{h_{1}k_{1}}{%
2\left\vert \pi _{1}\right\vert }$ in the subranges (a), (b), and (c) and $%
\tau _{2}=ih_{2}/c_{2}\approx \frac{h_{2}k_{2}}{2\left\vert \pi
_{2}\right\vert }$ in the whole range (\ref{4.4}). In these subranges we
have for $\left\vert \tau _{2}\right\vert $%
\begin{eqnarray}
\mathrm{(a)} &&\;1\leq \tau _{2}^{-1}<\left[ 1+\left( \sqrt{h_{2}}%
g_{2}\right) ^{-1}\right] ,  \notag \\
\mathrm{(b)} &&\;\left[ 1+\left( \sqrt{h_{2}}g_{2}\right) ^{-1}\right] <\tau
_{2}^{-1}<\left( 1+\varepsilon k_{2}/k_{1}\right) ,  \notag \\
\mathrm{(c)} &&\;\left( 1+\varepsilon k_{2}/k_{1}\right) <\tau _{2}^{-1}<%
\left[ 1+k_{2}/k_{1}\left( 1-1/g_{1}\right) \right] ,  \notag \\
\mathrm{(d)} &&\;\left[ 1+k_{2}/k_{1}\left( 1-1/g_{1}\right) \right] <\tau
_{2}^{-1}\lesssim \left( 1+k_{2}/k_{1}\right) .  \label{4.6}
\end{eqnarray}%
We see that $\tau _{1}-1\rightarrow 0$ and $\tau _{2}-1\rightarrow 0$ in the
range (a), while $\left\vert \tau _{1}-1\right\vert \sim 1$ in the range
(c), and $\left\vert \tau _{2}-1\right\vert \sim 1$ in the ranges (c) and
(d). In the range (b) these quantities vary from their values in the ranges
(a) and (c).

In the range (a) we can use the asymptotic expression of the confluent
hypergeometric function given by Eq.~(\ref{A1}) in Appendix \ref{Ap2}. Using
Eqs.~(\ref{A7}), (\ref{A8}), and (\ref{A9}) obtained in Appendix \ref{Ap2},
we finally find the leading term as%
\begin{equation}
N_{n}^{\mathrm{cr}}=e^{-\pi \lambda }\left[ 1+O\left( \left\vert \mathcal{Z}%
_{1}\right\vert \right) \right] ,  \label{4.7}
\end{equation}%
for fermions and bosons, where $\max \left\vert \mathcal{Z}_{1}\right\vert
\lesssim g_{2}^{-1}$ . In the range (c), we use the asymptotic expression of
the confluent hypergeometric function given by Eq.~(\ref{A10}) in Appendix %
\ref{Ap2}. Then we find that%
\begin{equation}
N_{n}^{\mathrm{cr}}=e^{-\pi \lambda }\left[ 1+O\left( \left\vert \mathcal{Z}%
_{1}\right\vert \right) ^{-1}+O\left( \left\vert \mathcal{Z}_{2}\right\vert
\right) ^{-1}\right] ,  \label{4.8}
\end{equation}%
where $\max \left\vert \mathcal{Z}_{1}\right\vert ^{-1}\lesssim \sqrt{%
g_{1}/h_{1}}$ and $\max \left\vert \mathcal{Z}_{2}\right\vert ^{-1}\lesssim
g_{2}^{-1}$. Using the asymptotic expression Eq.~(\ref{A1}) and taking into
account Eq.~(\ref{4.7}) and (\ref{4.8}), we can estimate that $N_{n}^{%
\mathrm{cr}}\sim e^{-\pi \lambda }$ in the range (b). In the range (d), the
confluent hypergeometric function $\Phi \left( a_{2},c_{2};ih_{2}\right) $
is approximated by Eq.~(\ref{A10a}) and the function $\Phi \left(
1-a_{1},2-c_{1};-ih_{1}\right) $ is approximated by Eq.~(\ref{A11}) given in
Appendix \ref{Ap2}. In this range the differential mean numbers in the
leading-order approximation are%
\begin{equation}
N_{n}^{\mathrm{cr}}\approx \frac{\exp \left[ -\frac{\pi }{k_{1}}\left(
\omega _{1}-\pi _{1}\right) \right] }{\sinh \left( 2\pi \omega
_{1}/k_{1}\right) }\times \left\{
\begin{array}{l}
\sinh \left[ \pi \left( \omega _{1}+\pi _{1}\right) /k_{1}\right] \ \mathrm{%
for\ fermions} \\
\cosh \left[ \pi \left( \omega _{1}+\pi _{1}\right) /k_{1}\right] \ \mathrm{%
for\ bosons}%
\end{array}%
\right. \,.  \label{4.9a}
\end{equation}%
It is clear that $N_{n}^{\mathrm{cr}}$ given by Eqs.~(\ref{4.9a}) tends to
Eq.~(\ref{4.8}), $N_{n}^{\mathrm{cr}}\rightarrow e^{-\pi \lambda }$, when $%
\pi _{1}\gg \pi _{\bot }$. Consequently, the forms (\ref{4.9a}) are valid in
the whole range (\ref{4.4}). Assuming $m/k_{1}\gg 1$, we see that \ values
of $N_{n}^{\mathrm{cr}}$ given by Eqs.~(\ref{4.9a}) are negligible in the
range $\pi _{1}\lesssim \pi _{\bot }$. Then we find for bosons and fermions
that significant value of $N_{n}^{\mathrm{cr}}$ is in the range $\pi _{\bot
}<\pi _{1}\leqslant eE/k_{1}$ and it has the form
\begin{equation}
N_{n}^{\mathrm{cr}}\approx \exp \left[ -\frac{2\pi }{k_{1}}\left( \omega
_{1}-\pi _{1}\right) \right] .  \label{4.10}
\end{equation}

Considering positive $p_{x}>0$, we can take into account that exact $N_{n}^{%
\mathrm{cr}}$ is invariant under the simultaneous exchange $%
k_{1}\leftrightarrows k_{2}$ and $\pi _{1}\leftrightarrows -\pi _{2}$. In
this case $\pi _{1}$ is positive and large, $\pi _{1}>eE/k_{1}$, while $\pi
_{2}$ varies from negative to positive values, $-eE/k_{2}<\pi _{2}<\infty $.
We find a significant contribution in the range
\begin{equation}
-h_{2}<2\pi _{2}/k_{2}<K_{2},  \label{4.11}
\end{equation}%
where $K_{2}$ is any given large number, $K_{2}\gg K_{\bot }$. In this
range, similarly to the case of the negative $p_{x}$, the differential mean
numbers in the leading-order approximation are%
\begin{equation}
N_{n}^{\mathrm{cr}}\approx \frac{\exp \left[ -\frac{\pi }{k_{2}}\left(
\omega _{2}+\pi _{2}\right) \right] }{\sinh \left( 2\pi \omega
_{2}/k_{2}\right) }\times \left\{
\begin{array}{c}
\sinh \left[ \pi \left( \omega _{2}-\pi _{2}\right) /k_{2}\right] \ \mathrm{%
for\ fermions} \\
\cosh \left( \pi \left( \omega _{2}-\pi _{2}\right) /k_{2}\right) \ \mathrm{%
for\ bosons}%
\end{array}%
\right. \,.  \label{4.12a}
\end{equation}%
Assuming $m/k_{2}\gg 1$, we find for bosons and fermions that significant
value of $N_{n}^{\mathrm{cr}}$ is in the range $-eE/k_{2}<\pi _{2}<-\pi
_{\bot }$ and it has a form%
\begin{equation}
N_{n}^{\mathrm{cr}}\approx \exp \left[ -\frac{2\pi }{k_{2}}\left( \omega
_{2}+\pi _{2}\right) \right] .  \label{4.13}
\end{equation}

Consequently, the quantity $N_{n}^{\mathrm{cr}}$ is almost constant over the
wide range of longitudinal momentum $p_{x}$ for any given $\lambda $
satisfying Eq.~(\ref{4.2}). When $h_{1},h_{2}\rightarrow \infty $, one
obtains the well-known result in a constant uniform electric field \cite%
{Nikis70a,Nikis70b},%
\begin{equation}
N_{n}^{\mathrm{cr}}\rightarrow N_{n}^{\mathrm{uni}}=e^{-\pi \lambda }.
\label{uni}
\end{equation}

\subsection{Total quantities\label{Ss4.2}}

In this subsection we estimate the total number $N^{\mathrm{cr}}$ of pairs
created by the peak electric field. To compute this number, one has to sum
the corresponding differential mean numbers $N_{n}^{\mathrm{cr}}$ over the
momenta $\mathbf{p}$ and, in the Fermi case, to{\Huge \ } sum over the spin
projections. Once $N_{n}^{\mathrm{cr}}$ does not depend on the spin
variables, the latter sum results in a multiplicative numerical factor $%
J_{(d)}=2^{\left[ d/2\right] -1}$ for fermions ($J_{\left( d\right) }=1$ for
bosons). Then replacing the sum over the momenta in Eq.~(\ref{TN}) by an
integral, the total number of pairs created from the vacuum takes the form%
\begin{equation}
N^{\mathrm{cr}}=\frac{V_{\left( d-1\right) }}{(2\pi )^{d-1}}J_{(d)}\int d%
\mathbf{p}N_{n}^{\mathrm{cr}}\,.  \label{asy5}
\end{equation}

Due to the structure of the coefficients $g\left( _{-}|^{+}\right) $
presented in section \ref{Ss2.3}, it is clear that a direct integration of
combinations of{\Huge \ }hypergeometric functions involved in the absolute
value of (\ref{gp}) and (\ref{gpKG}) is overcomplicated. Nevertheless the
analysis presented in section \ref{Ss4.1}\ reveals that the dominant
contributions for particle creation by a slowly varying field occurs in the
ranges of large kinetic momenta, whose differential quantities have the
asymptotic forms (\ref{4.10}) for $p_{x}<0$ and (\ref{4.13}) for $p_{x}>0$.
Therefore, one may represent the total number (\ref{asy5}) as%
\begin{eqnarray}
&&N^{\mathrm{cr}}=V_{\left( d-1\right) }n^{\mathrm{cr}}\,,\ \ n^{\mathrm{cr}%
}=\frac{J_{(d)}}{(2\pi )^{d-1}}\int_{\sqrt{\lambda }<K_{\bot }}d\mathbf{p}%
_{\bot }I_{\mathbf{p}_{\bot }},\ \ I_{\mathbf{p}_{\bot }}=I_{\mathbf{p}%
_{\bot }}^{\left( 1\right) }+I_{\mathbf{p}_{\bot }}^{\left( 2\right) },
\notag \\
&&I_{\mathbf{p}_{\bot }}^{\left( 1\right) }=\int_{-\infty }^{0}dp_{x}N_{n}^{%
\mathrm{cr}}\approx \int_{\pi _{\perp }}^{eE/k_{1}}d\pi _{1}\exp \left[ -%
\frac{2\pi }{k_{1}}\left( \omega _{1}-\pi _{1}\right) \right] \,,  \notag \\
&&I_{\mathbf{p}_{\bot }}^{\left( 2\right) }=\int_{0}^{\infty }dp_{x}N_{n}^{%
\mathrm{cr}}\approx \int_{\pi _{\perp }}^{eE/k_{2}}d\left\vert \pi
_{2}\right\vert \exp \left[ -\frac{2\pi }{k_{2}}\left( \omega
_{2}-\left\vert \pi _{2}\right\vert \right) \right] \,.  \label{tot2}
\end{eqnarray}

Using the change of the variables%
\begin{equation*}
s=\frac{2}{k_{1}\lambda }\left( \omega _{1}-\pi _{1}\right) \,,
\end{equation*}%
and neglecting exponentially small contributions, we represent the quantity $%
I_{\mathbf{p}_{\bot }}^{\left( 1\right) }$ as%
\begin{equation}
I_{\mathbf{p}_{\bot }}^{\left( 1\right) }\approx \int_{1}^{\infty }\frac{ds}{%
s}\omega _{1}e^{-\pi \lambda s}\,.  \label{4.15}
\end{equation}%
Similarly, using the change of variables%
\begin{equation*}
s=\frac{2}{k_{2}\lambda }\left( \omega _{2}-\left\vert \pi _{2}\right\vert
\right) \,,
\end{equation*}%
we represent the quantity $I_{\mathbf{p}_{\bot }}^{\left( 2\right) }$ as%
\begin{equation}
I_{\mathbf{p}_{\bot }}^{\left( 2\right) }\approx \int_{1}^{\infty }\frac{ds}{%
s}\omega _{2}e^{-\pi \lambda s}\,.  \label{4.16}
\end{equation}%
The leading contributions for both integrals (\ref{4.15}) and (\ref{4.16})
come from the range near $s\rightarrow 1$, where $\omega _{1}$ and $\omega
_{2}$ are approximately given by,%
\begin{equation*}
\omega _{1}\approx \frac{eE}{sk_{1}}\,,\ \ \omega _{2}\approx \frac{eE}{%
sk_{2}}\,.
\end{equation*}%
Consequently the leading term in $I_{\mathbf{p}_{\bot }}$ (\ref{tot2}) takes
the following final form,%
\begin{equation}
I_{\mathbf{p}_{\bot }}\approx \left( \frac{eE}{k_{1}}+\frac{eE}{k_{2}}%
\right) \int_{1}^{\infty }\frac{ds}{s^{2}}e^{-\pi \lambda s}=eE\left( \frac{1%
}{k_{1}}+\frac{1}{k_{2}}\right) e^{-\pi \lambda }G\left( 1,\pi \lambda
\right) ,  \label{4.17}
\end{equation}%
where%
\begin{equation}
G\left( \alpha ,x\right) =\int_{1}^{\infty }\frac{ds}{s^{\alpha +1}}%
e^{-x\left( s-1\right) }=e^{x}x^{\alpha }\Gamma \left( -\alpha ,x\right) ,
\label{4.19a}
\end{equation}%
and $\Gamma \left( -\alpha ,x\right) $ is the incomplete gamma function.

Neglecting the exponentially small contribution, one can represent the integral
over $\mathbf{p}_{\bot }$ in Eq.~(\ref{tot2}) (where $I_{\mathbf{p}_{\bot }}$
is given by Eq.~(\ref{4.17})) as
\begin{equation*}
\int_{\sqrt{\lambda }<K_{\bot }}d\mathbf{p}_{\bot }I_{\mathbf{p}_{\bot
}}\approx \int_{\sqrt{\lambda }<\infty }d\mathbf{p}_{\bot }I_{\mathbf{p}%
_{\bot }}.
\end{equation*}%
Then calculating the Gaussian integral,%
\begin{equation}
\int d\mathbf{p}_{\bot }\exp \left( -\pi s\frac{\mathbf{p}_{\bot }^{2}}{eE}%
\right) =\left( \frac{eE}{s}\right) ^{d/2-1},  \label{4.18}
\end{equation}%
we find%
\begin{equation}
n^{\mathrm{cr}}=r^{\mathrm{cr}}\left( \frac{1}{k_{1}}+\frac{1}{k_{2}}\right)
G\left( \frac{d}{2},\pi \frac{m^{2}}{eE}\right) ,\;\;r^{\mathrm{cr}}=\frac{%
J_{(d)}\left( eE\right) ^{d/2}}{(2\pi )^{d-1}}\exp \left\{ -\pi \frac{m^{2}}{%
eE}\right\} .  \label{4.19}
\end{equation}

Using the considerations presented above, one can perform the summation
(integration) in Eq.~(\ref{vacprob}) to obtain the vacuum-to-vacuum
probability $P_{v}$,%
\begin{eqnarray}
&&P_{v}=\exp \left( -\mu N^{\mathrm{cr}}\right) ,\;\;\mu =\sum_{l=0}^{\infty
}\frac{(-1)^{(1-\kappa )l/2}\epsilon _{l+1}}{(l+1)^{d/2}}\exp \left( -l\pi
\frac{m^{2}}{eE}\right) \;,  \notag \\
&&\epsilon _{l}=G\left( \frac{d}{2},l\pi \frac{m^{2}}{eE}\right) \left[
G\left( \frac{d}{2},\pi \frac{m^{2}}{eE}\right) \right] ^{-1}.  \label{4.20}
\end{eqnarray}

These results allow us to establish an immediate comparison with the
one-parameter regularizations of the constant field, namely the $T$-constant
and Sauter-like electric fields \cite{GavGit96}.\textsf{\ }We note that in
all these cases the quantity\ is quasiconstant over the wide range of the
longitudinal momentum $p_{x}$\ for any given\emph{\ }$\lambda ,$ i.e.,\emph{%
\ }$N_{n}^{\mathrm{cr}}\sim e^{-\pi \lambda }$\emph{. }Pair creation effects
in such fields are proportional to increments of longitudinal kinetic
momentum, $\Delta U=e\left\vert A_{x}\left( +\infty \right) -A_{x}\left(
-\infty \right) \right\vert $, which are%
\begin{eqnarray}
\Delta U_{\mathrm{p}} &=&eE\left( k_{1}^{-1}+k_{2}^{-1}\right) \;\mathrm{%
for\;peak\ field,}  \notag \\
\Delta U_{\mathrm{T}} &=&eET\;\;\mathrm{for\;T}\text{\textrm{-}}\mathrm{%
const\ field,}  \notag \\
\Delta U_{\mathrm{S}} &=&2eET_{\mathrm{S}}\;\;\mathrm{for\;Sauter}\text{%
\textrm{-}}\mathrm{like\ field}\text{.}  \label{4.21}
\end{eqnarray}%
This fact allows one to compare pair creation effects in such fields. Using
the quantities  introduced, we can represent the densities $n^{\mathrm{cr}}$
as follows:
\begin{eqnarray}
n^{\mathrm{cr}} &=&r^{\mathrm{cr}}\frac{\Delta U_{\mathrm{p}}}{eE}G\left(
\frac{d}{2},\pi \frac{m^{2}}{eE}\right) ,\ \mathrm{for\;peak\ field,}  \notag
\\
n^{\mathrm{cr}} &=&r^{\mathrm{cr}}\frac{\Delta U_{\mathrm{T}}}{eE}\;\;%
\mathrm{for\;T}\text{\textrm{-}}\mathrm{constant\ field},  \notag \\
n^{\mathrm{cr}} &=&r^{\mathrm{cr}}\frac{\Delta U_{\mathrm{S}}}{2eE}\delta \;%
\mathrm{for\;Sauter}\text{\textrm{-}}\mathrm{like\ field},  \label{4.23}
\end{eqnarray}%
where%
\begin{equation*}
\delta =\int_{0}^{\infty }dtt^{-1/2}(t+1)^{-\left( d-2\right) /2}\exp \left(
-t\pi \frac{m^{2}}{eE}\right) =\sqrt{\pi }\Psi \left( \frac{1}{2},-\frac{d-2%
}{2};\pi \frac{m^{2}}{eE}\right) ,
\end{equation*}%
and $\Psi \left( a,b;x\right) $ is the confluent hypergeometric function
\cite{BatE53}.

Thus, for a given magnitude of the electric field $E$ one can compare the
pair creation effects in fields with equal increment of the longitudinal
kinetic momentum, or one can determine such increments of the longitudinal
kinetic momenta,\emph{\ }for which particle creation effects are the same.
In Eq.(\ref{4.21}) $T$ is the time duration of the $T$-constant field.
Equating the densities $n^{\mathrm{cr}}$ for Sauter-like  field and 
for the  peak field to the density $n^{\mathrm{cr}}$ for the $T$-constant field, we find an effective duration time $T_{eff}$ in both cases,%
\begin{eqnarray}
T_{eff} &=&T_{\mathrm{S}}\delta \;\mathrm{for\;Sauter}\text{\textrm{-}}%
\mathrm{like\ field},  \notag \\
T_{eff} &=&\left( k_{1}^{-1}+k_{2}^{-1}\right) G\left( \frac{d}{2},\pi \frac{%
m^{2}}{eE}\right) \;\mathrm{for\;the\ peak\ field}.  \label{4.24}
\end{eqnarray}%
By the definition $T_{eff}=T$ for the $T$-constant field. One can say that
the Sauter-like and the peak electric fields with the same $T_{eff}=T$ are
equivalent to the $T$-constant field in pair production.

If the electric field $E$ is weak, $m^{2}/eE\gg $ $1$, one can use
asymptotic expressions for the $\Psi $-function and the incomplete gamma
function. Thus, we obtain
\begin{equation}
G\left( \frac{d}{2},\pi \frac{m^{2}}{eE}\right) \approx \frac{eE}{\pi m^{2}}%
,\ \ \delta \approx \sqrt{eE}/m\ .  \label{4.26}
\end{equation}%
If the electric field $E$ is strong enough, $m^{2}/eE\ll 1$, it follows from
a corresponding representation for the $\Psi $-function, see Ref.~\cite%
{BatE53}, that its leading term does not depend on the dimensionless
parameter $m^{2}/eE$ and reads%
\begin{equation}
\Psi \left( \frac{1}{2},-\frac{d-2}{2};\pi \frac{m^{2}}{eE}\right) \approx
\Gamma \left( d/2\right) /\Gamma \left( d/2+1/2\right) .  \label{asy12}
\end{equation}%
Then, for example, $\delta \approx \pi /2$ if $d=3$ and $\delta \approx 4/3$
if $d=4$. The leading term of $G$-function, which is given by Eq. (\ref%
{4.19a}), does not depend on the parameter $m^{2}/eE$ either,%
\begin{equation}
G\left( \frac{d}{2},\pi \frac{m^{2}}{eE}\right) \approx \frac{2}{d}.
\label{4.27}
\end{equation}

It is clear that there is a time range where Sauter-like and the peak
electric fields coincide with a $T$-constant field. Out of this range both
these fields have an exponential behavior and can be compared. Assuming $%
k_{1}\sim k_{2}$, we have%
\begin{eqnarray}
\dot{U} &\approx &eEe^{-2\left\vert t\right\vert /T_{\mathrm{S}}}\;\mathrm{if%
}\;\left\vert t\right\vert /T_{\mathrm{S}}\gg 1\;\mathrm{for\;Sauter}\text{%
\textrm{-}}\mathrm{like\ field},  \notag \\
\dot{U} &\approx &eEe^{-k_{1}\left\vert t\right\vert }\;\mathrm{if}%
\;k_{1}\left\vert t\right\vert \gg 1\;\mathrm{for\;peak\ field}.
\label{4.28}
\end{eqnarray}%
If the field is weak, $m^{2}/eE\gg $ $1$, we see that
\begin{equation*}
k_{1}T_{\mathrm{S}}=\frac{2\sqrt{eE}}{\pi m}\ll 1,
\end{equation*}%
that is, the peak electric field switches on and off much more slowly than the
Sauter-like field. If the field is strong, $m^{2}/eE\ll 1$, this
dimensionless parameter turns to unity,%
\begin{equation*}
k_{1}T_{\mathrm{S}}=\left\{
\begin{array}{l}
8/\left( 3\pi \right) \;\;\mathrm{if}\;d=3 \\
3/4\;\;\mathrm{if}\;d=4%
\end{array}%
\right. .
\end{equation*}%
In this case, the peak electric field switches on and off not much slowly
than the Sauter-like field.

Another global{\Huge \ }quantity is the vacuum-to-vacuum transition
probability $P_{v}$. It is given by Eq.~(\ref{4.20}) for the peak field and
has the form similar to that for the $T$-constant and the Sauter-like fields
with the corresponding $N^{\mathrm{cr}}$, and%
\begin{eqnarray}
\epsilon _{l} &=&\epsilon _{l}^{\mathrm{T}}=1\;\mathrm{for\;T}\text{\textrm{-%
}}\mathrm{constant\ field},  \notag \\
\epsilon _{l} &=&\epsilon _{l}^{\mathrm{S}}=\delta ^{-1}\sqrt{\pi }\Psi
\left( \frac{1}{2},-\frac{d-2}{2};l\pi \frac{m^{2}}{eE}\right) \;\mathrm{%
for\;Sauter}\text{\textrm{-}}\mathrm{like\ field}.  \label{4.29}
\end{eqnarray}%
If the field is weak, $m^{2}/eE\gg  1$, then $\;\epsilon _{l}^{\mathrm{S}%
}\approx l^{-1/2}$ for the Sauter-like field and $\epsilon _{l}\approx
l^{-1} $ for the peak field. Then $\mu \approx 1$ for both fields and we see
that the identification with $T_{eff}=T$, given by Eq.~(\ref{4.24}), is the
same as the one extracted from the comparison of total densities $n^{\mathrm{%
cr}}$. In the case of a strong field, $m^{2}/eE\ll 1$, all the terms with
different $\epsilon _{l}^{\mathrm{S}}$ and $\epsilon _{l}$ contribute
significantly to the sum in Eq.~(\ref{4.20}) if $l\pi m^{2}/eE\sim 1$, and
the $\mu $ quantities differ essentially from the case of the $T$-constant
field. However, for a very strong field, $l\pi m^{2}/eE\ll 1$, the leading
contribution for $\epsilon _{l}$ has a quite simple form $\epsilon _{l}^{%
\mathrm{S}}\approx \epsilon _{l}\approx 1$. In this case the quantities $\mu
$ are the same for all these fields, namely%
\begin{equation*}
\mu \approx \sum_{l=0}^{\infty }\frac{(-1)^{(1-\kappa )l/2}}{(l+1)^{d/2}},
\end{equation*}%
and the identification with $T_{eff}=T$ is the same as the one extracted
from the comparison of the total densities $n^{\mathrm{cr}}$.

It is clear that different total quantities, such as the total number of
created pairs and the vacuum-to-vacuum transition probability {\Huge \ }%
discussed above, in the general case lead to different identifications with $%
T_{eff}=T$. We believe that some of these quantities are more adequate for
such an identification. In this connection, it should be noted that in
small-gradient fields, the total vacuum mean values, such as mean electric
currents and the mean energy-momentum tensor, are usually of interest; see,
e.g. Refs. \cite{GavGit08,GavGit08b,GavGitY12}. These total quantities are represented
by corresponding sums of differential numbers of created particles.
Therefore, relations between the total numbers and parameters $\Delta U_{%
\mathrm{p}}$, $\Delta U_{\mathrm{T}}$, and $\Delta U_{\mathrm{S}}$ derived
above are also important. Such relations derived from the vacuum-to-vacuum
transition probability $P_{v}$\ are interesting in semiclassical approaches
based on Schwinger's technics \cite{Sch51}. We recall that the semiclassical%
{\LARGE \ }approaches work in the case of weak external fields $m^{2}/eE\gg $%
\emph{\ }$1$. It should be noted that in the case of a strong field when the
semiclassical approach is not applicable, the probability $P_{v}$\ has no
direct relation to vacuum mean values of the above discussed physical
quantities.

\section{Configurations with sharp fields\label{S5}}

\subsection{Short pulse field\label{Ss5.1}}

Choosing certain parameters of the peak field, one can obtain electric
fields that exist only for a short time in a vicinity of the time instant $%
t=0$. The latter fields switch on and/or switch off{\Huge \ }%
\textquotedblleft abruptly\textquotedblright\ near the time instant $t=0$.
Let us consider large parameters $k_{1}$, $k_{2}\rightarrow \infty $ with a
fixed ratio $k_{1}/k_{2}$. The corresponding asymptotic potentials, $U\left(
+\infty \right) =eEk_{2}^{-1}$ and $U\left( -\infty \right) =-eEk_{1}^{-1}$
define finite increments of the longitudinal kinetic momenta $\Delta U_{1}$
and $\Delta U_{2}$ for increasing and decreasing parts, respectively, \ \ \
\ \ ,%
\begin{equation}
\Delta U_{1}=U\left( 0\right) -U\left( -\infty \right)
=eEk_{1}^{-1},\;\;\Delta U_{2}=U\left( +\infty \right) -U\left( 0\right)
=eEk_{2}^{-1}.  \label{5.1}
\end{equation}%
Such a case corresponds to a very short pulse of the electric field. At the
same time this configuration imitates well enough a $t$-electric rectangular
potential step (it is an analog of the Klein step, which is an $x$-electric
rectangular step; see Ref.~\cite{GavGit15}) and coincides with it as $k_{1}$%
, $k_{2}\rightarrow \infty $. Thus, these field configurations can be
considered as regularizations of rectangular step. We assume that
sufficiently large $k_{1}$ and $k_{2}$ satisfy the following inequalities:
\begin{equation}
\Delta U_{1}/k_{1}\ll 1,\;\;\Delta U_{2}/k_{2}\ll 1,\;\;\max \left( \omega
_{1}/k_{1},\omega _{2}/k_{2}\right) \ll 1  \label{5.2}
\end{equation}%
for any given $\pi _{\perp }$ and $\pi _{1,2}=p_{x}-U\left( \mp \infty
\right) $. In this case the confluent hypergeometric function can be
approximated by the first two terms in Eq. (\ref{chf}), which are $\Phi
\left( a,c;\eta \right) $, $c_{j}\approx 1$, and $\ \ a_{j}\approx \left(
1+\chi \right) /2$. Then for fermions, we obtain the result
\begin{equation}
N_{n}^{\mathrm{cr}}=\frac{\left( \omega _{1}+\pi _{1}\right) \left( \Delta
U_{2}+\Delta U_{1}+\omega _{2}-\omega _{1}\right) ^{2}}{4\omega _{1}\omega
_{2}\left( \omega _{2}-\pi _{2}\right) }  \label{5.3}
\end{equation}%
which does not depend on $k_{1,2}$. For bosons, we obtain%
\begin{equation}
N_{n}^{\mathrm{cr}}=\frac{\left( \omega _{2}-\omega _{1}\right) ^{2}}{%
4\omega _{1}\omega _{2}}.  \label{5.4}
\end{equation}

In contrast to the Fermi case, where $N_{n}^{\mathrm{cr}}\leq 1$, in the
Bose case, the differential numbers $N_{n}^{\mathrm{cr}}$ are unbounded in
two ranges of the longitudinal kinetic momenta, in the range where $\omega
_{1}/\omega _{2}\rightarrow \infty $ and in the range where $\omega
_{2}/\omega _{1}\rightarrow \infty $. In these ranges they are%
\begin{equation}
N_{n}^{\mathrm{cr}}\approx \frac{1}{4}\max \left\{ \omega _{1}/\omega
_{2},\omega _{2}/\omega _{1}\right\} .  \label{5.5}
\end{equation}

If $k_{1}=k_{2}$ (in this case $\Delta U_{2}=\Delta U_{1}=\Delta U/2$), we
can compare the above results with the results of the regularization of
rectangular steps by the Sauter-like potential \cite{AdoGavGit15},\ obtained
for a small $T_{\mathrm{S}}\rightarrow 0$\ and constant $\Delta U=2eET_{%
\mathrm{S}}$\ under the conditions\emph{\ }$\Delta UT_{\mathrm{S}}\ll 1$%
\emph{\ }and\emph{\ }$\max \left\{ T_{\mathrm{S}}\omega _{1},T_{\mathrm{S}%
}\omega _{2}\right\} \ll 1$. We see that both regularizations are in
agreement for fermions under the condition $\left\vert \omega _{2}-\omega
_{1}\right\vert \ll \Delta U$ , and for bosons under the condition $\left(
\omega _{2}-\omega _{1}\right) ^{2}\gg \left( \Delta U\right) ^{4}T_{\mathrm{%
S}}^{2}/4$, which is the general condition for applying the Sauter-like
potential for the regularization of rectangular step\emph{\ }for bosons.

\subsection{Exponentially decaying field\label{Ss5.2}}

In the examples, considered above, the pick field switches on and off
relatively smooth. Here we are going to consider a different essentially
asymmetric configuration of the peak field, when for example, the field
switches abruptly on at $t=0$, that is, $k_{1}$ is sufficiently large, while
the value of parameter $k_{2}>0$ remains arbitrary and includes the case of
a smooth switching off. Note that due to the invariance of the mean numbers $%
N_{n}^{\mathrm{cr}}$ under the simultaneous change $k_{1}\leftrightarrows
k_{2}$ and $\pi _{1}\leftrightarrows -\pi _{2}$, one can easily transform
this situation to the case with a large $k_{2}$ and arbitrary $k_{1}>0$.

Let us assume that a sufficiently large $k_{1}$ satisfies the inequalities
\begin{equation}
\Delta U_{1}/k_{1}\ll 1,\;\;\omega _{1}/k_{1}\ll 1.  \label{5.6}
\end{equation}%
Then Eqs.~(\ref{4.0}) and (\ref{4b}) can be reduced to the following form%
\begin{equation}
\left\vert \Delta \right\vert ^{2}\approx \left\vert \Delta _{\mathrm{ap}%
}\right\vert ^{2}=e^{i\pi \nu _{2}}\left. \left\vert \left[ \chi \Delta
U_{1}+\omega _{2}-\omega _{1}+k_{2}h_{2}\left( -\frac{1}{2}+\frac{d}{d\eta
_{2}}\right) \right] \Phi \left( a_{2},c_{2};\eta _{2}\right) \right\vert
^{2}\right\vert _{t=0}\,.  \label{5.7}
\end{equation}%
Under the condition%
\begin{equation}
-2p_{x}\Delta U_{1}\ll \pi _{\perp }^{2}+p_{x}^{2},  \label{5.8}
\end{equation}%
one can disregard the term $\chi \Delta U_{1}$ in Eq.~(\ref{5.7}) and write
approximately $\pi _{1}\approx p_{x}$. Thus, $\omega _{1}\approx \sqrt{%
p_{x}^{2}+\pi _{\perp }^{2}}$. In this approximation, leading terms do not
contain $\Delta U_{1}$, so that we obtain
\begin{equation}
N_{n}^{\mathrm{cr}}\approx \left\{
\begin{array}{l}
\left\vert \mathcal{C}\Delta _{\mathrm{ap}}\right\vert ^{2}\ \mathrm{for\
fermions} \\
\left\vert \mathcal{C}_{\mathrm{sc}}\left. \Delta _{\mathrm{ap}}\right\vert
_{\chi =0}\right\vert ^{2}\ \mathrm{for\ bosons}%
\end{array}%
\right. \,.  \label{5.9}
\end{equation}%
In fact, differential mean numbers obtained in these approximations are the
same as in the so-called exponentially decaying electric field, given by the
potential%
\begin{equation}
A_{x}^{\mathrm{ed}}\left( t\right) =E\left\{
\begin{array}{l}
0\,,\ \ t\in \mathrm{I}\, \\
k_{2}^{-1}\left( e^{-k_{2}t}-1\right) \,,\ \ t\in \mathrm{II}%
\end{array}%
\right. \,.  \label{com1}
\end{equation}%
The effect of pair creation in the exponentially decaying electric field was
studied previously by us in Ref. \cite{AdoGavGit14}. Note that the pair
creation due to an exponentially decaying background has been studied in de
Sitter spacetime and for the constant electric field in two dimensional de
Sitter spacetime; see, e.g., \cite{AndMot14,StStHue16} and references
therein. Under condition (\ref{5.8}), the results presented by Eqs.~(\ref%
{5.9}) for arbitrary $k_{2}>0$ are in agreement with ones obtained in Ref.
\cite{AdoGavGit14}.

Let us consider the most asymmetric case when Eqs.~(\ref{5.9}) hold and when
the increment of the longitudinal kinetic momentum due to exponentially
decaying electric field is sufficiently large ($k_{2}$ are sufficiently
small),%
\begin{equation}
h_{2}=2\Delta U_{2}/k_{2}\gg \max \left( 1,m^{2}/eE\right) \,.  \label{5.11}
\end{equation}

As it was noted in Section \ref{Ss4.1}, in this case only the range of fixed
$\pi _{\perp }$ is essential and we assume that the inequality (\ref{4.2})
holds. In the case under consideration $K_{\bot }$ is any given number
satisfying the condition%
\begin{equation}
h_{2}\gg K_{\bot }^{2}\gg \max \left( 1,m^{2}/eE\right) \,.  \label{5.12}
\end{equation}

It should be noted that the distribution $N_{n}^{\mathrm{cr}}$, given by
Eqs.~(\ref{5.9}) for this most asymmetric case\emph{\ }coincides with the
one obtained in our recent work \cite{AdoGavGit14}, where the exponentially
decreasing field was considered. However, the detailed study of this
distribution was not performed there. In the following, we study how this
distribution depends on the parameters $p_{x}$ and $\pi _{\perp }$.

In the case of large negative $p_{x}$, $p_{x}<0$ and $\left\vert
p_{x}\right\vert /\sqrt{eE}>K_{\bot }$, using appropriate asymptotic
expressions of the confluent hypergeometric function, given in Appendix \ref%
{Ap2}, one can conclude that numbers $N_{n}^{\mathrm{cr}}$ are negligibly
small both for fermions and bosons. The same holds true for very large
positive $p_{x}$, such that $2\pi _{2}/k_{2}>K_{2}$, where $K_{2}$ is any
given large number, $K_{2}\gg K_{\bot }$. We see that $N_{n}^{\mathrm{cr}}$
are nonzero\ only in the range%
\begin{equation}
-K_{\bot }<p_{x}/\sqrt{eE},\;\;2\pi _{2}/k_{2}<K_{2}.  \label{5.13}
\end{equation}%
This range can be divided in three subranges,%
\begin{eqnarray}
\mathrm{(a)} &&\;\left( 1-\varepsilon \right) h_{2}\leq -2\pi
_{2}/k_{2}<\left( 1+\varepsilon \right) h_{2},  \notag \\
\mathrm{(b)} &&\;h_{2}/g_{1}<-2\pi _{2}/k_{2}<\left( 1-\varepsilon \right)
h_{2},  \notag \\
\mathrm{(c)} &&\;-K_{2}<-2\pi _{2}/k_{2}<h_{2}/g_{1},  \label{5.14}
\end{eqnarray}%
where $g_{1}$ and $\varepsilon $ are any given numbers satisfying the
conditions $g_{1}$ $\gg 1$ and $\varepsilon \ll 1$. We assume that $%
\varepsilon \sqrt{h_{2}}\gg 1$. Note that%
\begin{equation*}
\tau _{2}=ih_{2}/c_{2}\approx \frac{h_{2}k_{2}}{2\left\vert \pi
_{2}\right\vert }
\end{equation*}%
in the ranges (a) and (b). Then in the ranges (a) and (b), $\tau _{2}$
varies from $1-\varepsilon $ to $g_{1}$. In the range (b), parameters $\eta
_{2}$ and $c_{2}$ are large with $a_{2}$ fixed and $\tau _{2}>1$. In this
case, using the asymptotic expression of the confluent hypergeometric
function given by Eq.~(\ref{A10}) in Appendix \ref{Ap2}, we find that%
\begin{equation}
N_{n}^{\mathrm{cr}}=\exp \left[ -\frac{2\pi }{k_{2}}\left( \omega _{2}+\pi
_{2}\right) \right] \left[ 1+O\left( \left\vert \mathcal{Z}_{2}\right\vert
^{-1}\right) \right]  \label{5.15}
\end{equation}%
both for fermions and bosons, where $\mathcal{Z}_{2}$ is given by Eq.~(\ref%
{A5}) in the Appendix \ref{Ap2}. We note that\textrm{\ }modulus $\left\vert
\mathcal{Z}_{2}\right\vert ^{-1}$ varies from $\left\vert \mathcal{Z}%
_{2}\right\vert ^{-1}\sim \left( \varepsilon \sqrt{h_{2}}\right) ^{-1}$ to $%
\left\vert \mathcal{Z}_{2}\right\vert ^{-1}\sim \left[ \left( g_{1}-1\right)
\sqrt{h_{2}}\right] ^{-1}$. Approximately, expression (\ref{5.15}) can be
written as%
\begin{equation}
N_{n}^{\mathrm{cr}}\approx \exp \left( -\frac{\pi \pi _{\perp }^{2}}{%
k_{2}\left\vert \pi _{2}\right\vert }\right) .  \label{5.16}
\end{equation}%
Note that $eE/g_{1}<k_{2}\left\vert \pi _{2}\right\vert <\left(
1-\varepsilon \right) eE$ in the range (b).

It is clear that the distribution $N_{n}^{\mathrm{cr}}$ given by Eq.~(\ref%
{5.16}) has the following limiting form:%
\begin{equation*}
N_{n}^{\mathrm{cr}}\rightarrow e^{-\pi \lambda }\ \ \mathrm{as\ \ }%
k_{2}\left\vert \pi _{2}\right\vert \rightarrow \left( 1-\varepsilon \right)
eE\ .
\end{equation*}%
In the range (a), we can use the asymptotic expression for the confluent
hypergeometric function given by Eq.~(\ref{A1}) in Appendix \ref{Ap2} to
verify that $N_{n}^{\mathrm{cr}}$ is finite and restricted, $N_{n}^{\mathrm{%
cr}}\lesssim e^{-\pi \lambda },$ both for fermions and bosons. Thus, we see
that the well-known distribution obtained by Nikishov \cite{Nikis70a} in a
constant uniform electric field is reproduced in an exponentially decaying
electric field in the range of a large increment of the longitudinal kinetic
momentum, $-\pi _{2}\sim eE/k_{2}$.

In the range (c), we can use the asymptotic expression of the confluent
hypergeometric function for large $h_{2}$ with fixed $a_{2}$ and $c_{2}$
given by Eq.~(\ref{A11}) in Appendix \ref{Ap2} to get the following result:
\begin{equation}
N_{n}^{\mathrm{cr}}\approx \frac{\exp \left[ -\frac{\pi }{k_{2}}\left(
\omega _{2}+\pi _{2}\right) \right] }{\sinh \left( 2\pi \omega
_{2}/k_{2}\right) }\times \left\{
\begin{array}{c}
\sinh \left[ \pi \left( \omega _{2}-\pi _{2}\right) /k_{2}\right] \ \mathrm{%
for\ fermions} \\
\cosh \left( \pi \left( \omega _{2}-\pi _{2}\right) /k_{2}\right) \ \mathrm{%
for\ bosons}%
\end{array}%
\right. \,  \label{5.17}
\end{equation}%
in the leading-order approximation. The same distribution was obtained for $%
p_{x}>0$ in a slowly varying field; see Eq.~(\ref{4.12a}).

For $m/k_{2}\gg 1$, distribution~(\ref{5.17}) has the form (\ref{5.15}),
which means that distribution (\ref{5.15}) holds in the range (c) as well.

Using the above considerations, we can estimate the total number $N^{\mathrm{%
cr}}$ (\ref{asy5}) of pairs created by an exponentially decaying electric
field. To this end, we represent the leading terms of integral (\ref{asy5})
as a sum of two contributions, one due to the range (a) and another due to
the ranges (b) and (c):
\begin{eqnarray}
&&N^{\mathrm{cr}}=V_{\left( d-1\right) }n^{\mathrm{cr}}\,,\ \ n^{\mathrm{cr}%
}=\frac{J_{(d)}}{(2\pi )^{d-1}}\int_{\sqrt{\lambda }<K_{\bot }}d\mathbf{p}%
_{\bot }I_{\mathbf{p}_{\bot }},\ \ I_{\mathbf{p}_{\bot }}=I_{\mathbf{p}%
_{\bot }}^{\left( 1\right) }+I_{\mathbf{p}_{\bot }}^{\left( 2\right) },
\notag \\
&&I_{\mathbf{p}_{\bot }}^{\left( 1\right) }=\int_{\pi _{2}\in \mathrm{(a)}%
}d\pi _{2}N_{n}^{\mathrm{cr}}\,,\ \ I_{\mathbf{p}_{\bot }}^{\left( 2\right)
}=\int_{\pi _{2}\in \mathrm{(b)\cup (c)}}d\pi _{2}N_{n}^{\mathrm{cr}}\,.
\label{5.18}
\end{eqnarray}%
Note that numbers $N_{n}^{\mathrm{cr}}$ given by Eqs.~(\ref{5.17}) are
negligibly small in the range $-\pi _{2}\lesssim \pi _{\bot }$. Then the
integral $I_{\mathbf{p}_{\bot }}^{\left( 2\right) }$ in Eq.~(\ref{5.18}) can
be taken from Eq.~(\ref{tot2}). Using the results of section \ref{Ss4.2}, we can
verify that the leading term in $I_{\mathbf{p}_{\bot }}^{\left( 2\right) }$
takes the following final form:%
\begin{equation}
I_{\mathbf{p}_{\bot }}^{\left( 2\right) }\approx \frac{eE}{k_{2}}%
\int_{1}^{\infty }\frac{ds}{s^{2}}e^{-\pi \lambda s}=\frac{eE}{k_{2}}e^{-\pi
\lambda }G\left( 1,\pi \lambda \right) ,  \label{5.19}
\end{equation}%
where $G\left( \alpha ,x\right) $ is given by Eq.~(\ref{4.19a}). The
integral $I_{\mathbf{p}_{\bot }}^{\left( 1\right) }$ in Eq.~(\ref{5.18}) is
of the order of $e^{-\pi \lambda }\varepsilon eE/k_{2}$, such that it is
relatively small comparing to integral (\ref{5.19}). Thus, the dominant
contribution is given by integral (\ref{5.19}), $I_{\mathbf{p}_{\bot
}}\approx I_{\mathbf{p}_{\bot }}^{\left( 2\right) }$. Then calculating the
Gaussian integral, we find%
\begin{equation}
n^{\mathrm{cr}}=\frac{r^{\mathrm{cr}}}{k_{2}}G\left( \frac{d}{2},\pi \frac{%
m^{2}}{eE}\right) ,  \label{5.20}
\end{equation}%
where $r^{\mathrm{cr}}$ is given by Eq.~(\ref{4.19}). We see that $n^{%
\mathrm{cr}}$ given by Eq.~(\ref{5.20}) is the $k_{2}$-dependent part of the
mean number density of pairs created in the slowly varying peak field~(\ref%
{4.19}). In the case of a strong field, $n^{\mathrm{cr}}$ given by Eq.~(\ref%
{5.20}) has the form obtained in Ref.\textrm{\cite{AdoGavGit14}.}

Finally, we can see that the vacuum-to-vacuum probability is%
\begin{equation}
P_{v}=\exp \left( -\mu N^{\mathrm{cr}}\right) ,  \label{5.21}
\end{equation}%
where $\mu $ is given by Eq.~(\ref{4.20}).

\section{Concluding remarks}

Using the strong-field QED, we consider for the first time particle creation
in the so-called peak electric field, which is a combination of two
exponential parts, one exponentially increasing and another
exponentially decreasing. This is an addition to the few previously known
exactly solvable cases, where one can perform nonperturbative calculations
of all the characteristics of particle creation process. For a
certain choice of parameters, the peak electric field produces a particle%
creation effect similar to that of the Sauter-like electric field,
or of{\Huge \ }the constant electric field, or of the electric pulse of
laser beams. Besides, by varying the peak field parameters we can change the
asymmetry rate so that the resulting field turns out to be effectively
equivalent to the exponentially decaying field. All these asymptotic regimes
are discussed in detail and a comparison with the pure asymptotically
decaying field is considered. Moreover, the results obtained allow one to
study how the effects of switching on and off together or separately affect the
particle creation processes. Changing the parameters of the peak electric field
we can adjust its form to a specific physical situation, in particular,
imitate field configurations characteristic for graphene, Weyl semimetals
and so on.

\section*{Acknowledgements}

The work of the authors{\large \ }was supported by a grant from the Russian
Science Foundation, Research Project No. 15-12-10009.

\appendix

\section{Some asymptotic expansions \label{Ap2}}

The asymptotic expression of the confluent hypergeometric function for large
$\eta $ and $c$ with fixed $a$ and $\tau =\eta /c\sim 1$ is given by
Eq.~(13.8.4) in \cite{DLMF} as%
\begin{eqnarray}
&&\Phi \left( a,c;\eta \right) \simeq c^{a/2}e^{\mathcal{Z}^{2}/4}F\left(
a,c;\tau \right) ,\ \ \mathcal{Z=-}\left( \tau -1\right) \mathcal{W}\left(
\tau \right) \sqrt{c},  \notag \\
&&F\left( a,c;\tau \right) =\tau \mathcal{W}^{1-a}D_{-a}\left( \mathcal{Z}%
\right) +\mathcal{R}D_{1-a}\left( \mathcal{Z}\right) ,  \notag \\
&&\mathcal{R}=\left( \mathcal{W}^{a}-\tau \mathcal{W}^{1-a}\right) /\mathcal{%
Z},\ \ \mathcal{W}\left( \tau \right) =\left[ 2\left( \tau -1-\ln \tau
\right) /\left( \tau -1\right) ^{2}\right] ^{1/2}  \label{A1}
\end{eqnarray}%
where $D_{-a}\left( \mathcal{Z}\right) $ is the Weber parabolic cylinder
function (WPCF) \cite{BatE53}. Using Eq.~(\ref{A1}) we present the functions
$y_{1}^{2}\left( \eta _{2}\right) $, $y_{2}^{1}\left( \eta _{1}\right) $ and
their derivatives at $t=0$ as%
\begin{eqnarray}
&&\left. y_{2}^{1}\left( \eta _{1}\right) \right\vert _{t=0}\simeq
e^{ih_{1}/2}\left( ih_{1}\right) ^{-\nu _{1}}\left( 2-c_{1}\right) ^{\left(
1-a_{1}\right) /2}e^{\mathcal{Z}_{1}^{2}/4}F\left( 1-a_{1},2-c_{1};\tau
_{1}\right) ,  \notag \\
&&\mathcal{Z}_{1}\mathcal{=}\mathcal{-}\left( \tau _{1}-1\right) \mathcal{W}%
\left( \tau _{1}\right) \sqrt{2-c_{1}},\ \ \tau _{1}=-ih_{1}/\left(
2-c_{1}\right) ,  \notag \\
&&\left. \frac{\partial y_{2}^{1}\left( \eta _{1}\right) }{\partial \eta _{1}%
}\right\vert _{t=0}\simeq e^{ih_{1}/2}\left( ih_{1}\right) ^{-\nu
_{1}}\left( 2-c_{1}\right) ^{\left( 1-a_{1}\right) /2}e^{\mathcal{Z}%
_{1}^{2}/4}\left[ -\frac{1}{2ih_{1}}-\frac{1}{2-c_{1}}\frac{\partial }{%
\partial \tau _{1}}\right] F\left( 1-a_{1},2-c_{1};\tau _{1}\right) ;  \notag
\\
&&\left. y_{1}^{2}\left( \eta _{2}\right) \right\vert _{t=0}\simeq
e^{-ih_{2}/2}\left( ih_{2}\right) ^{\nu _{2}}c_{2}^{a_{2}/2}e^{\mathcal{Z}%
_{2}^{2}/4}F\left( a_{2},c_{2};\tau _{2}\right) ,  \notag \\
&&\mathcal{Z}_{2}\mathcal{=}\mathcal{-}\left( \tau _{2}-1\right) \mathcal{W}%
\left( \tau _{2}\right) \sqrt{c_{2}},\ \ \tau _{2}=ih_{2}/c_{2},  \notag \\
&&\left. \frac{\partial y_{1}^{2}\left( \eta _{2}\right) }{\partial \eta _{2}%
}\right\vert _{t=0}\simeq e^{-ih_{2}/2}\left( ih_{2}\right) ^{\nu
_{2}}c_{2}^{a_{2}/2}e^{\mathcal{Z}_{2}^{2}/4}\left[ -\frac{1}{2ih_{2}}+\frac{%
1}{c_{2}}\frac{\partial }{\partial \tau _{2}}\right] F\left(
a_{2},c_{2};\tau _{2}\right) .  \label{A5}
\end{eqnarray}

Assuming $\tau -1\rightarrow 0$, one has%
\begin{eqnarray*}
&&\mathcal{W}^{1-a}\approx 1+\frac{a-1}{3}\left( \tau -1\right) ,\ \
\mathcal{R}\approx \frac{2\left( a+1\right) }{3\sqrt{c}},\ \ \mathcal{%
Z\approx -}\left( \tau -1\right) \sqrt{c}, \\
&&\frac{\partial F\left( a,c;\tau \right) }{\partial \tau }\approx \frac{2+a%
}{3}D_{-a}\left( \mathcal{Z}\right) +\frac{\partial D_{-a}\left( \mathcal{Z}%
\right) }{\partial \tau }+\mathcal{R}\frac{\partial D_{1-a}\left( \mathcal{Z}%
\right) }{\partial \tau }.
\end{eqnarray*}%
Expanding WPCFs near $\mathcal{Z}=0$, one obtains that%
\begin{eqnarray}
&&\frac{\partial F\left( a,c;\tau \right) }{\partial \tau }\approx -\sqrt{%
\eta }D_{-a}^{\prime }\left( 0\right) +d\left( a,c;\tau \right) ,  \notag \\
&&F\left( a,c;\tau \right) \approx D_{-a}\left( 0\right) +c^{-1/2}f\left(
a,c;\tau \right) ,  \label{A2}
\end{eqnarray}%
in the next-to-leading approximation at $\mathcal{Z}\rightarrow 0$, where $%
f\left( a,c;\tau \right) $ and $d\left( a,c;\tau \right) $ are the
next-to-leading terms,%
\begin{eqnarray}
&&d\left( a,c;\tau \right) =\left[ \frac{2+a}{3}+\left( a-\frac{1}{2}\right)
\left( \tau -1\right) c\right] D_{-a}\left( 0\right) +\frac{2}{3}\left(
a-2\right) D_{1-a}^{\prime }\left( 0\right) ,  \notag \\
&&f\left( a,c;\tau \right) =-\left( \tau -1\right) cD_{-a}^{\prime }\left(
0\right) +\frac{2}{3}\left( 2-a\right) D_{1-a}\left( 0\right) ,  \label{A3}
\end{eqnarray}%
and%
\begin{equation}
D_{-a}\left( 0\right) =\frac{2^{-a/2}\sqrt{\pi }}{\Gamma \left( \frac{a+1}{2}%
\right) },\ \ D_{-a}^{\prime }\left( 0\right) =\frac{2^{\left( 1-a\right) /2}%
\sqrt{\pi }}{\Gamma \left( \frac{a}{2}\right) },  \label{A4}
\end{equation}%
where $\Gamma (z)$ is the Euler gamma function. We find under condition (\ref%
{4.1}) that%
\begin{eqnarray}
&&\omega _{1,2}\approx \left\vert \pi _{1,2}\right\vert \left( 1+\lambda
/h_{1,2}\right) ,\ \ a_{1,2}\approx \left( 1+\chi \right) /2+i\lambda /2,
\notag \\
&&2-c_{1}\approx 1-i\left( \lambda +\frac{2\pi _{1}}{k_{1}}\right) ,\ \
c_{2}\approx 1+i\left( \lambda -\frac{2\pi _{2}}{k_{2}}\right) ,  \notag \\
&&\tau _{1}-1\approx -\frac{1}{h_{1}}\left( i+\lambda +\frac{2p_{x}}{k_{1}}%
\right) ,\ \ \tau _{2}-1\approx \frac{1}{h_{2}}\left( i-\lambda +\frac{2p_{x}%
}{k_{2}}\right) .  \label{A6}
\end{eqnarray}%
Using Eqs.~(\ref{A5}) and (\ref{A6}) we represent Eq.~(\ref{4.0}) in the form%
\begin{eqnarray}
&&N_{n}^{\mathrm{cr}}=B^{2}\left\vert \tilde{\Delta}\right\vert ^{2},\ \
\tilde{\Delta}\simeq k_{1}h_{1}F\left( a_{2},c_{2};\tau _{2}\right) \left[ -%
\frac{1}{2ih_{1}}-\frac{1}{2-c_{1}}\frac{\partial }{\partial \tau _{1}}%
\right] F\left( 1-a_{1},2-c_{1};\tau _{1}\right)  \notag \\
&&+k_{2}h_{2}F\left( 1-a_{1},2-c_{1};\tau _{1}\right) \left[ -\frac{1}{%
2ih_{2}}+\frac{1}{c_{2}}\frac{\partial }{\partial \tau _{2}}\right] F\left(
a_{2},c_{2};\tau _{2}\right) ,  \notag \\
&&B^{2}=e^{-\pi \lambda /2}\left( 2eE\right) ^{-1}\left[ 1+O\left(
h_{1}^{-1}\right) +O\left( h_{2}^{-1}\right) \right] .  \label{A7}
\end{eqnarray}%
Taking into account Eqs.~(\ref{A2}) we obtain%
\begin{eqnarray}
&&\left\vert \tilde{\Delta}\right\vert ^{2}\approx \left\vert \tilde{\Delta}%
_{0}\right\vert ^{2}+2\mathrm{Re}\left[ \tilde{\Delta}_{0}\left( k_{1}\delta
_{1}+k_{2}\delta _{2}\right) \right] ,  \notag \\
&&\tilde{\Delta}_{0}=\sqrt{2eE}\left[ e^{i\pi /4}D_{-a_{2}}\left( 0\right)
D_{a_{1}-1}^{\prime }\left( 0\right) -e^{-i\pi /4}D_{-a_{2}}^{\prime }\left(
0\right) D_{a_{1}-1}\left( 0\right) \right] ,  \notag \\
&&\delta _{1}=iD_{-a_{2}}\left( 0\right) \left[ \frac{1}{2}D_{a_{1}-1}\left(
0\right) -d\left( 1-a_{1},-ih_{1};\tau _{1}\right) \right]
-D_{-a_{2}}^{\prime }\left( 0\right) f\left( 1-a_{1},-ih_{1};\tau
_{1}\right) ,  \notag \\
&&\delta _{2}=iD_{a_{1}-1}\left[ \frac{1}{2}D_{-a_{2}}\left( 0\right)
-d\left( a_{2},ih_{2};\tau _{2}\right) \right] +D_{a_{1}-1}^{\prime }f\left(
a_{2},ih_{2};\tau _{2}\right) ,  \label{A8}
\end{eqnarray}%
where functions$\ d$ and $f$ are given by Eq.~(\ref{A3}). Assuming $\chi =1$
for fermions and $\chi =0$ for bosons, and using the relations of the Euler
gamma function we find that%
\begin{equation}
\tilde{\Delta}_{0}=\sqrt{2eE}\exp \left( i\pi /2+i\pi \chi /4\right) e^{-\pi
\lambda /4}.  \label{A9}
\end{equation}

Assuming $\left\vert \tau -1\right\vert \sim 1$, one can use the asymptotic
expansions of WPCFs in Eq.~(\ref{A1}), e.g., see \cite{BatE53,DLMF}. Note
that $\arg \left( \mathcal{Z}\right) \approx \frac{1}{2}\arg \left( c\right)
$ if $1-\tau >0$. Then one finds that%
\begin{equation}
\Phi \left( a,c;\eta \right) =\left( 1-\tau \right) ^{-a}\left[ 1+O\left(
\left\vert \mathcal{Z}\right\vert ^{-1}\right) \right] \ \ \mathrm{if}\
\;1-\tau >0.  \label{A10a}
\end{equation}%
In the case of $1-\tau <0$, one has
\begin{equation*}
\arg \left( \mathcal{Z}\right) \approx \left\{
\begin{array}{c}
\frac{1}{2}\arg \left( c\right) +\pi \ \ \mathrm{if}\ \ \arg \left( c\right)
<0 \\
\frac{1}{2}\arg \left( c\right) -\pi \ \ \mathrm{if}\ \ \arg \left( c\right)
>0%
\end{array}%
\right. .
\end{equation*}%
Then one obtains finally that%
\begin{equation}
\Phi \left( a,c;\eta \right) =\left\{
\begin{array}{l}
\left( \tau -1\right) ^{-a}e^{-i\pi a}\left[ 1+O\left( \left\vert \mathcal{Z}%
\right\vert ^{-1}\right) \right] \ \ \mathrm{if}\ \;\arg \left( c\right) <0
\\
\left( \tau -1\right) ^{-a}e^{i\pi a}\left[ 1+O\left( \left\vert \mathcal{Z}%
\right\vert ^{-1}\right) \right] \ \ \mathrm{if}\ \;\arg \left( c\right) >0%
\end{array}%
\right. .  \label{A10}
\end{equation}

The asymptotic expression of the confluent hypergeometric function $\Phi
\left( a,c;\pm ih\right) $ for large real $h$ with fixed $a$ and $c$ is
given by Eq.~(6.13.1(2)) in \cite{BatE53} as%
\begin{equation}
\Phi \left( a,c;\pm ih\right) =\frac{\Gamma \left( c\right) }{\Gamma \left(
c-a\right) }e^{\pm i\pi a/2}h^{-a}+\frac{\Gamma \left( c\right) }{\Gamma
\left( a\right) }e^{\pm ih}\left( e^{\pm i\pi /2}h\right) ^{a-c}.
\label{A11}
\end{equation}

\end{document}